\documentclass[12pt]{article} 

\usepackage{epsfig}
\usepackage{graphicx}
\usepackage{amsmath}
\usepackage{amssymb}
\usepackage{url}

\begin{document}

\title{The flyby anomaly: A multivariate analysis approach}

\author{L. Acedo\thanks{E-mail: luiacrod@imm.upv.es}\\
Instituto Universitario de Matem\'atica Multidisciplinar,\\
Building 8G, $2^{\mathrm{o}}$ Floor, Camino de Vera,\\
Universitat Polit$\grave{\mbox{e}}$cnica de Val$\grave{\mbox{e}}$ncia,\\
Valencia, Spain\\
}

\maketitle

\begin{abstract}
The flyby anomaly is the unexpected variation of the asymptotic post-encounter velocity of
a spacecraft with respect to the pre-encounter velocity as it performs a slingshot manoeuvre.
This effect has been detected in, at least, six flybys of the Earth but it has not appeared in
other recent flybys. In order to find a pattern in these, apparently contradictory, data several
phenomenological formulas have been proposed but all have failed to predict a new result in
agreement with the observations. In this paper we use a multivariate dimensional analysis approach
to propose a fitting of the data in terms of the local parameters at perigee, as it would occur if
this anomaly comes from an unknown fifth force with latitude dependence. Under this assumption, we estimate the range of this force around $300$ km.
\end{abstract}


%
\section{Introduction}
\label{intro}

In the last quarter-century there have been important advances in the high-accuracy measurements
of spacecraft, planets and moons's ephemerides in the Solar system through Doppler and laser ranging \cite{Dickey1994,Williams2004}.
On parallel with these developments, faster computers and improved numerical methods have allowed to
test the predictions of standard orbit determination models with a precision never achieved before \cite{DE431}.
As a consequence of these improvements several discrepancies among the theoretical models and the 
observations have been disclosed \cite{IorioIJMPDReview}. Among them we can enumerate: (i) The Pioneer anomaly
\cite{TuryshevReview,PioneerPRL} (ii) The flyby
anomaly \cite{Anderson2008} (iii) The anomalous secular increase of the astronomical unit \cite{Krasinsky} (iv) An unexplained secular increase in the eccentricity of the orbit of the Moon \cite{Iorio2011MNRAS,Iorio2011AJ} (v) The Faint Young Sun Paradox
\cite{Iorio2013Galaxies} and other \cite{IorioIJMPDReview}.
In many of these anomalies we cannot exclude that further research should render them statistically not significant. But this is certainly not the case for the Pioneer anomaly (and also for the flyby anomaly)
which has also revealed very clearly in the fitting of orbital data.

The Pioneer 10 and Pioneer 11 are the first spacecraft within the context of the ``Grand Tour'' program
whose objective was to send robotic spacecraft to all the planets of outer Solar system from Jupiter to Neptune and, more recently, also Pluto \cite{Butrica}. These spacecraft are provided with a wide antenna designed for
sending downlink signals and receiving uplinks from the Earth. Thanks to the monitoring of the
spacecraft throughout the years it was found  an anomalous constant drift of the redshifted signal \cite{TuryshevReview,Anderson2002}, to
be interpreted as an acceleration directed, approximately, towards the Sun with magnitude
$a_P=(8.74\pm 1.33) \times 10^{-8}$ cm$/$s$^2$. Many conventional and unconventional proposals were
proposed to explain away this anomaly to no avail until the whole dataset for the mission was retrieved \cite{TuryshevReview}.
The analysis of this data showed that this acceleration diminishes with time with the same time-scale that the thermal recoil force arising from the anisotropic emission of thermal radiation off the spacecraft \cite{PioneerPRL,Rievers2011}. The spacecraft heat is diffusing from the radioisotope thermoelectric generators filled
with Plutonium 238, whose half-life is $87.74$ years. This correlation eventually lead to a complete
explanation of the anomaly in terms of a thermal radiation effect. For this reason, it is useful
to look for this kind of systematic behaviour of correlations in the case of the flyby anomaly as well,
because it could also provide a clue about its origin.

On December, 8th, 1990 the Galileo spacecraft performed a flyby of the Earth at a minimum altitude at perigee of $960$ km. After processing Doppler tracking data for obtaining a fit to the trajectory, NASA
engineers found that the post-encounter and pre-encounter trajectories cannot be accomodated into the
same model and that a small unexplainable residue remained in the Doppler data corresponding to
a difference of the post-encounter and pre-encounter asymptotic velocities of $3.92$ mm$/$s \cite{Anderson2008,LPDSolarSystem}. Later on, a second flyby was performed on December, 8th, 1992 in which a total residual velocity decrease of $-8$ mm$/$s was found. However, it has been estimated that $-3.4$ mm$/$s should correspond to atmospheric
friction because in this flyby the spacecraft crossed through the middle of the thermosphere. Anyway, it was concluded that an anomalous velocity decrease of $-4.6$ mm$/$s remains to be explained. The maximum
anomaly was found in the NEAR flyby of January, 23th, 1998 in which an increase of $13.46$ mm$/$s has been unexplained to date. Similar anomalies were also detected at the Cassini and Rosetta flybys but they were not detected (or were below the threshold of measurement errors) in the Messenger flyby \cite{Anderson2008}, the second and third Rosetta flybys \cite{Jouannic} and, more recently, in the Juno flyby of October, 9th, 2013 \cite{Thompson}.

Nowadays, Doppler ranging and Delta-Differential one-way ranging have achieved an impressive nanosecond accuracy. It is 
estimated that random effects, such as clock instability or fluctuations in the atmosphere, could account for delay errors
up to $0.053$ nanoseconds \cite{DeltaDOR}.  Consequently, the results for the flyby anomaly are sufficiently precise to be attributed only to measurement errors.

In their seminal paper about this problem, Anderson et al. \cite{Anderson2008} proposed a phenomological formula which fitted rather well the six flybys whose data was available at the time. According to Anderson and his team the anomalous velocity variation is given by:

\begin{equation}
\label{AndersonF}
\Delta V_\infty = V_\infty K \left( \cos \delta_i-\cos \delta_o \right)\; ,
\end{equation}

where $V_\infty$ is the osculating asymptotic velocity at perigee, $\delta_i$, $\delta_o$ are the declinations for the incoming and outgoing velocity vectors and $K$ is a constant. These authors
also ventured to postulate that $K$ is related to the quotient of the tangential velocity of the Earth at the Equator and the speed of light as follows:

\begin{equation}
\label{Kpar}
K=\displaystyle\frac{2 \, \Omega \, R_E}{c}=3.099 \times 10^{-6} \mbox{ s$^{-1}$}\; .
\end{equation}

Here $\Omega=2 \pi/86400$ s$^{-1}$ is the angular velocity for the Earth's rotation around its axis, $R_E=6371$ km is the average Earth's radius and $c$ is the speed of light in vacuum. The fit provided
by Eq. (\ref{AndersonF}) was good for the six flybys studied in their paper but it fails to
provide a prediction for the null results of the Rosetta II and III, and the Juno flybys. By proposing
a model such as that in Eqs. (\ref{AndersonF}) and (\ref{Kpar}) these authors are also hinting at an 
explanation in terms of an unknown axisymmetric interaction arising from the rotating Earth, which they
called an enhanced Lense-Thirring effect \cite{Anderson2008}. 

The point of view for an ignored classical effect has also been discussed in detail \cite{LPDSolarSystem}. In particular, general relativistic effects \cite{IorioSRE2009,Hackmann}, Lorentz's charge acceleration \cite{Atchison} and thermal radiation \cite{Rievers2011}. Other studies have looked for an explanation beyond standard physics: an halo of dark matter around the Earth \cite{Adler2010,Adler2011}, extensions
of general relativity \cite{Hafele,Acedo2014,Acedo2015,Pinheiro2014,Pinheiro2016} or phenomenologica formulas \cite{Jouannic,Busack} , alternative to that of Eq. (\ref{AndersonF}). But none of these approaches have provided a satisfactory fitting of all the data available and, even more, they have not explained the null results for the anomalies of subsequent flybys (with the exception of the work of Pinheiro \cite{Pinheiro2016}). The difficulty to find
a pattern in the anomalies, allowing them to be fitted by a single phenomenological formula, comes from  
two facts: (i) the anomaly corresponds either to an increase (Galileo I, NEAR and Rosetta I flybys) or
a decrease (Galileo II and Cassini) of the asymptotic velocity of the spacecraft (ii) in some cases
no anomalous increase or decrease of the asymptotic velocity has been detected (Messenger, Rosetta II, III and Juno flybys).
On the other hand, this null effect has been anticipated by Pinheiro \cite{Pinheiro2016} for flybys in the prograde
direction in the context of a topological torsion current model. But, the problem persists if we assume that the origin
of the anomaly is a velocity-independent field of force as assumed in this paper.

Moreover, point (ii) suggest the irreproducibility of the results in point (i) as announced by Anderson et al. \cite{Anderson2008}. This is event more evident for the Juno and NEAR flybys whose perigee's altitudes were
similar. But, in this paper we will pursue the case for a flyby anomaly originating for an unknown
field (or fifth force) as it has been proposed before. Our approach will be a general multivariate
statistical analysis to test if the data for the nine flybys, whose results have been reported until
now, can be encompassed by an expression in terms of the local parameters at perigee (as expected in the
effect arises from an interaction with the rotating Earth). These parameters are the geocentric latitude of the perigee, $\phi$, the orbital inclination, $I$, and the perigee's altitude, $h$. We will show that
expressions  for $\Delta V_\infty/V_\infty$, in terms of these parameters, with a coefficient of determination $R^2 > 0.9$ are possible. Anyway, a very fast decay of the predicted ratio $\Delta V_\infty/V_\infty$ with altitude is necessary for obtaining a reasonably good fit. This implies that, in case the flyby anomalies originate in an unknown force field, this field extends only to the Earth's exosphere and it is almost undetectable at higher altitudes.

The paper is organized as follows: In section \ref{data} we analyze the available data and their correlations including the azimuthal and polar components of the velocity at perigee. A multivariate fitting model is discussed in section \ref{fitting} and we apply it to the recent Juno flyby of Jupiter. 
The papers ends with section \ref{discuss} with a brief discussion, and the implications of our model for the study of the anomalies, in the near future.

\section{Flyby data}
\label{data}

In this section we will collect the data for the nine flybys enumerated in the previous section.
Most of these parameters can be retrieved directly from the NASA's ephemeris website: such as the altitude at perigee and the latitude of the point on the surface of the Earth lying at the vertical
of the perigee \cite{Horizons}. Notice that if we use equatorial celestial coordinates the declination of the 
spacecraft coincides with the geocentric latitude \cite{Vallado}. The asymptotic velocity, $V_\infty$, in Anderson et
al. \cite{Anderson2008} phenomenological model corresponds to the asymptotic value of the velocity for the osculating
hyperbolic at perigee. This idealized orbit is defined in terms of parameters at perigee as the 
orbit that the spacecraft would follow if all the perturbations were switched off (mainly the perturbations from the Sun and the Moon). These perturbations give rise to variations of $V_\infty$ over the whole data interval of the flyby manoeuvre in the range of a few meters per second. Anyway, the
average of the incoming velocities a day before and after the closest approach is a good approximation
for $V_\infty$ if our purpose is to obtain a phenomenological formula for the flyby anomalies.

Alternatively, we can obtain the parameters for the osculating orbit at perigee by considering the 
celestial coordinates and the altitude of the perigee and another point of the trajectory a minute
(or a few minutes) after the closest approach. If we denote by $r_t$ the distance to the center of
the Earth at $t=1$ min and by $r_P$ the distance at perigee, we get that the orbital eccentricity
$\epsilon$ at perigee is found as the solution of the system \cite{Danby,Burns}

\begin{eqnarray}
\label{osculating}
r_t &=& r_P \displaystyle\frac{\epsilon \cosh \eta-1}{\epsilon-1} \; ,\\
\noalign{\smallskip}
t &=& \sqrt{\displaystyle\frac{r_P^3}{\mu}}\left(\epsilon \sinh \eta-\eta\right) \; ,
\end{eqnarray}

where $\eta$ is the eccentric anomaly for the osculating orbit at time $t$ and $\mu= G M_E =398600.4$ km$^3/$s$^2$ is the product of the gravitational constant and the mass of the Earth. This way we find also the semi-major axis, $a$, of the osculating keplerian orbit from the relation:

\begin{equation}
\label{adef}
r_P = a \left( 1 - \epsilon \right)\; .
\end{equation}

From here we can define the time-scale, $T=\sqrt{-a^3/\mu}$. The osculating hyperbolic asymptotic velocity at perigee is then defined as $V_\infty = -a/T$ and the velocity at perigee is given by:

\begin{equation}
\label{Vperigee}
V_{\mbox{perigee}}= V_\infty \sqrt{\displaystyle\frac{\epsilon+1}{\epsilon-1}} \; .
\end{equation}
The direction of this vector can also be found from an orbital frame of reference defined by the
inclination vector, the unit position vector at perigee and the tangential vector obtained as the
cross product of the other two vectors \cite{Acedo2014,Acedo2015}.

Another important parameter is the orbital inclination that we found from the cross product of 
the position vectors at perigee and time $t=1$ min as follows:
\begin{equation}
\label{inclination}
\cos I=\left(\displaystyle\frac{\mathbf{r}_P \times \mathbf{r}_t}{\left\vert \mathbf{r}_P \times \mathbf{r}_t \right \vert} \right)_z\; .
\end{equation}
Here we divide by the magnitude of the cross product to obtain a unit vector and $z$ denotes the third
component of the cross product vector. The definition in Eq. (\ref{inclination}) is ambiguous because
for any angle $I$ we can also choose $180^\circ -I$. This ambiguity is solved by defining the inclination vector according to the right-hand rule.

By following these criteria we obtain the orbital inclination in Table \ref{tab1} and also the rest
of parameters we need for our model: the perigee's altitude and latitude, the anomalous velocity
increase and the asymptotic velocity. In this table we also list the azimuthal and the polar components 
of the velocity at perigee obtained by projecting the velocity vector at perigee for the osculating 
orbit onto the unit azimuthal vector and the  unit polar vector at that point.

Some correlations are manifested from simple inspection of the data in Table \ref{tab1}. In particular
we find that for retrograde orbits (those in which the spacecraft moves opposite to the rotation of
the Earth) the anomalous velocity increase is positive, being negative for prograde orbits (for the Cassini flyby). Anyway, this correlation is broken by the Galileo II flyby in which a retrograde
orbit was accompanied by an anomalous decrease of $V_\infty$ \cite{Acedo2016}.

%
%

%

\begin{table*}[tb]
\small
\caption{Parameters for the nine flybys mentioned in the main text: orbital inclination, $I$, and geocentric latitude, $\phi$, in sexagesimal degrees, perigee's altitute, $h$, in km, asymptotic velocity , $V_\infty$, in km$/$s, anomalous velocity increase $\Delta V_\infty$ in mm$/$s, azimuthal, $V_a$, and polar, $V_p$, components of the velocity at perigee in km$/$s.} 
\label{tab1}
\resizebox{\textwidth}{!}{
\begin{tabular}{lcccccccc}
\hline  
  Spacecraft & Date & Inclination & Latitude  & h (km) & $V_\infty$  & $\Delta V_\infty$ 
  & $V_a$ (km$/$s) & $V_p$ (km$/$s) \\
  \noalign{\smallskip}
  Galileo I & 12/8/1990 & $142.9^\circ$ & $25.2^\circ$ & 960 & 8.949 & 3.92 & -11.993 & 6.707\\
  \noalign{\smallskip}
  Galileo II & 12/8/1992 & $138.7^\circ$ & $-33.8^\circ$ & 303 & 8.877 & -4.60 & -12.729 & 6.018\\
  \noalign{\smallskip}
  NEAR & 1/23/1998 & $108.0^\circ$ & $33.0^\circ$ & 539 & 6.851 & 13.46 & -4.694 & 11.843\\
  \noalign{\smallskip}
  Cassini   & 8/18/1999 & $25.4^\circ$ & $-23.5^\circ$ & 1175 & 16.010 & -2 & 18.740 & -3.279\\
  \noalign{\smallskip}
  Rosetta I & 3/4/2005 & $144.9^\circ$ & $20.20^\circ$ & 1956 & 3.863 & 1.8 & -9.170 & 5.154\\
  \noalign{\smallskip}
  Messenger & 8/2/2005 & $43.05^\circ$ & $46.95^\circ$ & 2347 & 4.056 & 0.02 & -10.387 & -0.353\\
 \noalign{\smallskip}
  Juno & 9/10/2013 & $47.13^\circ$ & $-33.39^\circ$ & 559 & 10.389 & 0 & 11.845 & -8.429\\ 
 \noalign{\smallskip}
  Rosetta II & 13/11/2007 & $25.08^\circ$ & $-64.76^\circ$ & 5322 & 5.064 & 0 & -12.463 & -1.356\\ 
 \noalign{\smallskip}
  Rosetta III & 13/11/2009 & $65.63^\circ$ & $-7.44^\circ$ & 2483 & 9.393 & 0 & -12.263 & -5.274\\
  \noalign{\smallskip}
\hline 
\end{tabular}}
%
\end{table*}

A better correlation appears among the latitude of the perigee and the sign of the anomalous
variation of the asymptotic velocity. The cases in which an anomalous decrease of the asymptotic velocity was found are the Galileo II and the Cassini flybys in which the vertical 
of the perigee was located in the southern hemisphere. It is also clear that the anomalies tend
to decrease, in absolute value, for higher perigee's altitudes and, in particular, there are null results reported for the Rosetta II and Rosetta III flybys. This is also evident for the EPOXI flybys of 2007, 2008 and 2009 performed at
even higher altitudes \cite{Jouannic}. But the most conflicting data is the
one corresponding to the Juno flyby of October, 9th, 2013: its altitude was similar to that of
the NEAR flyby with a geocentric latitude at perigee almost coincident with that of the Galileo II
flyby. If we notice that in these two flybys the anomalies with the largest magnitudes were found, it is
difficult to justify, within a model searching for a new local interaction, the reason for the null
result in the Juno flyby. However, the Juno shares with the Messenger flyby that the orbit was inclined
almost $45^\circ$ with respect to the equatorial plane. In the next section, we will propose a fitting
multivariate formula taking into account this fact. 

\section{Multivariate fitting analysis}
\label{fitting}

In this section we will propose a phenomenological fitting formula for the flyby anomalies listed
in Table \ref{tab1}. Later on, we will apply this formula to the recent Juno flyby of Jupiter on
August, 27th, 2016 in order to estimate the magnitude of the anomalies that could be found in
these flybys after processing the data for the trajectories.

The formula for the Earth flybys will take into account the following correlations found in the
qualitative analysis in the previous section and also dimensional analysis consistency:

\begin{itemize}
\item The anomalous velocity change, $\Delta V_\infty$, should be proportional to the asymptotic velocity for the osculating orbit at perigee, $V_\infty$.
\item The anomalies decrease with the altitude of the perigee. So, the expression should include a
power of $R_E/(R_E+h)$, $R_E=6371$ km being the average radius of the Earth.
\item The anomaly cancels for orbits with a $45^\circ$ inclination angle at perigee. For this reason, we will include a
factor $\cos 2 I$ in the phenomenological formula for the anomalous velocity increase.
A possible field configuration giving rise to this cancellation of the anomalous velocity variation, for this particular
orbit inclination, is obtained by a assuming two components of the field (equal in magnitude) along the polar and the 
azimuthal directions ($\vert \mathbf{F}_\theta \vert=\vert \mathbf{F}_\phi\vert$. If the spacecraft moves along $\mathbf{F}_\theta$ and opposite to $\mathbf{F}_\phi$, or viceversa, we have:
\begin{equation}
\label{DeltaV45}
\Delta V(t)=\displaystyle\sqrt{ \left(\mathbf{V}_p+\delta \mathbf{V}_\theta(t)\right)^2+\left(\mathbf{V}_p-\delta \mathbf{V}_\phi(t)\right)^2}-V_p\; ,
\end{equation}
where $\mathbf{V}_p$ is the velocity vector at perigee, $\Delta V(t)$ is the anomalous velocity variation at time $t$ after
perigee and $\delta \mathbf{V}_\theta(t)$, $\delta \mathbf{V}_\phi(t)$ are the velocity perturbations induced by the polar and
azimuthal components of the field, respectively. If $\mathbf{V}_p$ is oriented at a $45^\circ$ inclination angle the scalar products in
Eq. (\ref{DeltaV45}) cancel out and we have:
\begin{equation}
\Delta V(t)\simeq \delta V \left(\displaystyle\frac{\delta V}{V_p}\right) \ll \delta V \; ,
\end{equation}
where $\delta V=\vert \delta \mathbf{V}_\theta(t)\vert=\vert \delta \mathbf{V}_\phi(t)\vert$. Then, Eq. (\ref{DeltaV45}) implies that the anomaly 
is almost undetectable in this particular configuration in contrast with flybys along the celestial parallels or meridians (in which case the variation $\Delta V(t)$ would not be suppressed by the small ratio $\delta V/V_p$).

\item A null result is also expected for orbits whose perigee is attained at the poles. This is predicted by Anderson et al. \cite{Anderson2008} phenomenological formula and it is a reasonable assumption for interactions
arising from the Earth's rotation. In our model this is achieved by including a term $\sin 2 \phi$, where $\phi$ is the latitude of the perigee. The latitude, $\phi$, is coincident with the declination, $\delta$, in the celestial equatorial coordinate system.
\item The anomaly is positive (negative) for latitudes $\phi > 0$ ($\phi < 0$), respectively. This condition and the previous one are verified by a term $\sin 2 \phi$. This condition suggests that the underlying force, originating the anomaly, has a quadrupolar distribution around the Earth.
\end{itemize}

Under these assumptions our proposal for a phenomenological formula relating $\Delta V_\infty$ and the
aforementioned parameters is:

\begin{equation}
\label{DVform}
\Delta V_\infty= \kappa\, V_\infty \left( \displaystyle\frac{R_E}{R_E+h} \right)^\beta\, \left\vert
\cos 2 I \right\vert^\gamma \, \sin 2 \phi\;.
\end{equation}

The fit to the data in Table \ref{tab1} gives us the coefficients listed in Table \ref{tab2}. The
corresponding coefficient of determination is $R^2=0.9429$, denoting that most of the variability of
the data can be explained in term of these three parameters, $h$, $I$ and $\phi$.
A consequence of Eq. (\ref{DVform}) is that the escape velocity of a spacecraft from a putative fifth force
field goes as $1/(R_E+h)^\beta$ in comparison with the $1/(R_E+h)^{1/2}$ dependence for standard Newtonian gravity. This
suggests that the fifth force decrease much faster (with the distance to the center of the Earth) than Newtonian gravity
or gravitomagnetic effects in General Relativity. Nonetheless, it is difficult to find a rationale for this distance
dependence at this stage of the research on the flyby anomaly but it could point towards the existence of a very low mass
boson mediating the interaction. One of these X bosons in the dark sector could have been already detected \cite{Feng}.
%
\begin{table*}[tb]
\caption{Parameters $\kappa$, $\gamma$ and $\delta$ obtained after fitting the flyby data in 
Table \protect\ref{tab1} to the expression in Eq. (\protect\ref{DVform}). The standard error and
the $t$-statistics ratio and $p$-values are also shown.} 
\label{tab2}
\begin{center}
\begin{tabular}{lcccc}
\hline  
Parameter & Estimated value & Standard Error & t-statistics & p-value \\
\noalign{\smallskip}
$\kappa$ & $16.2247\times 10^{-6}$ & $7.9816\times 10^{-6}$ & $2.0327$ & $0.0883$ \\
\noalign{\smallskip}
$\beta$ & $21.8739$ & $5.4244$ & $4.0324$ & $0.0068$ \\
\noalign{\smallskip}
$\gamma$ & $1.1537$ & $0.1922$  & $6.00187$ & $0.00096$\\
\noalign{\smallskip}
\hline 
\end{tabular}
\end{center}
\end{table*}

We also notice that the $p$-values are statistically significant for the exponents $\beta$ and $\gamma$
($p < 0.05$) but it is also close to be signficant for the prefactor $\kappa$. We can also notice
that $\kappa$ can be written as follows:

\begin{equation}
\label{kappaeq}
\kappa= \kappa' \, \displaystyle\frac{\Omega\, R_E}{c}\; , \mbox{ with $\kappa'=10.47 \pm 5.15$}\; .
\end{equation}

The predictions of this model in Eq. (\ref{DVform}) and the standard error of these predictions
are displayed in Table \ref{tab3}.

\begin{table}[tb]
\caption{Prediction and standard error vs the observed anomalies for the nine flybys in 
Table \protect\ref{tab1}. Velocity anomalies are measured in mm$/$s.} 
\label{tab3}
\begin{center}
\begin{tabular}{lccc}
\hline  
Flyby & Prediction & Error & Observation \\
\noalign{\smallskip}
Galileo I & $1.157$ & $1.546$ & $3.92$ \\
\noalign{\smallskip}
Galileo II & $-4.529$ & $2.015$ & $-4.6$ \\
\noalign{\smallskip}
NEAR & $13.457$ & $2.066$  & $13.46$\\
\noalign{\smallskip}
Cassini & $-2.759$ & $1.992$ & $-2$ \\
\noalign{\smallskip}
Rosetta & $0.033$ & $1.463$ & $1.8$ \\
\noalign{\smallskip}
Messenger & $0.003$ & $1.463$ & $0.02$ \\
\noalign{\smallskip}
Juno & $-1.228$ &  $1.558$ &  $0$ \\
\noalign{\smallskip}
Rosetta II & $0.000$ & $1.463$ & $0$ \\
\noalign{\smallskip}
Rosetta III & $-0.018$ & $1.463$ & $0$ \\
\noalign{\smallskip}
\hline 
\end{tabular}
\end{center}
\end{table}

We see that according to Table \ref{tab3} only the predictions for the Galileo II, NEAR and Cassini
flybys are different from zero within their error bars. The other, including the Messenger, Juno and the second and third Rosetta flybys are compatible with a null result in the search for a flyby anomaly as it has been reported.

On the other hand, the predictions of Anderson et al. phenomenological formula in Eq. (\ref{AndersonF}) 
can be obtained for Juno by taking into account and incoming declination angle, $\delta_i=-14.308^\circ$, and an outgoing angle, $\delta_o=39.409^\circ$. This gives $\Delta V_\infty=6.3355$ mm$/$s in disagreement with the data obtained after the orbir reconstruction.
Similarly, we have the predictions $\Delta V_\infty=0.523$ mm$/$s and $\Delta V_\infty=1.099$ mm$/$s for the Rosetta II and the Rosetta III flybys, respectively. These are smaller and could still agree with
the predictions of the alternative formula we have proposed in this paper.

The main problem for a further understanding of this puzzling anomaly is the lack of data. But, at present the Juno mission is being carried out with a total of 36 close flybys of Jupiter in the mission program. The first of these flybys took place on August, 27th, 2016 with the Juno spacecraft approaching
at a minimum distance around $4200$ km over the the top clouds of the giant planet (with radius $R_J=71492$ km). The tangential velocity at periapsis was estimated as $57.77$ km$/$s. We can also obtain the orbital inclination at 
periapsis and the declination from the Horizons web-interface proceeding as we discussed in the previous section. This way we find: $\delta_P=28.56^\circ$ and $I=90.31^\circ$, but Jupiter's rotation axis is only slightly tilted with respect to the celestial equator so we can assume that the declination is similar to the latitude with respect to Jupiter's equator ($\phi_P \simeq \delta_P$). To apply our 
phenomenological formula we should also take into account that Jupiter's rotational period around its
axis is $9.925$ hours and this changes Anderson's ratio $K$ to $K_J=8.115\times 10^{-5}$ s$^{-1}$ and, consequently, also the value of the parameter $\kappa$ in Eq. (\ref{kappaeq}). We must also mention
that Juno's trajectory is now elliptical because it is trapped by Jupiter's gravitational field, so it is meaningless to define the osculating hyperbolic orbit at periapsis or $V_\infty$. Anyway, we
should use $V_P$ instead to estimate the perturbation of the putative unknown field generated by Jupiter's rotation on Juno as it passed through the periapsis. This way we obtain a prediction of
$\Delta V=6$ m$/$s for the post-encounter velocity with respect to the pre-encounter velocity. This difference is three orders of magnitude larger than the typical anomaly for Earth flybys and it should
be detected in the orbit reconstruction of the Juno flybys of Jupiter. 

If this is not the case, and no anomalies are found at these close flybys of Juno over Jupiter, we should cast a serious doubt on the reality of the phenomenon and the statistical significance of the previous results on Earth flybys.

\section{Conclusions and discussion}
\label{discuss}

In this paper we have considered the data for the flyby anomaly as reported in nine gravity assist manoeuvres from 1990 to 2013. In five of these flybys significant anomalies were reported after trying
to find a single fit for the post-encounter and pre-encounter Doppler tracked trajectories. This motivated a team at JPL lead by Anderson to propose a daring hypothesis on the connection of this phenomenon with an unkown field produced by the rotating Earth and to find a phenomenological formula
which fitted reasonably well the data known until 2008 \cite{Anderson2008}. Other  authors have also tried to fit the 
anomalies \cite{Acedo2014,Cahill,McCulloch,Pinheiro2014,Pinheiro2016}. On the other hand, other flybys of the Earth
by spacecraft developed at ESA, such as the second and third Rosetta flybys, and NASA, the Juno flyby of the Earth in 2013, have failed to detect any anomaly in the difference among the post-encounter and
the pre-encounter velocities \cite{Thompson}. The impasse in this problem is enhanced by the fact that the Juno flyby
was performed at altitudes and latitudes similar to previous flybys in which the anomaly has been detected. Apart from conventional explanations: such as atmospheric drag \cite{LPDSolarSystem}, ocean and solid tides \cite{AcedoMNRAS}, electric charge or magnetic moment of the spacecraft \cite{Atchison}, or relativistic effects \cite{IorioSRE2009,Hackmann}, there have also been
some attempts for modified models of gravity in order to make some sense of the flyby anomaly data \cite{Hafele,Acedo2015}.

In this paper we have suggested an alternative phenomenological formula to that of Anderson et al. under
the hypothesis that the effect arises as a short-ranged field from the rotating Earth which decays very
fast with altitude and whose effect depends on the orbital orientation at perigee. In the fitting formula, the null results for the Juno and Messenger flybys are obtained, in part, as a result of its particular inclination, around $45^\circ$,. over the celestial equator. We also consider a latitude dependence in which the sign of the anomaly depends on the hemisphere flybyed by the spacecraft and
we implement also the condition of symmetry under which the anomaly is null for flybys whose perigee is
attained at the Earth poles.

The resulting formula in Eq. (\ref{DVform}) is adjusted using multivariate analysis techniques and
a fitting with a $R^2 > 0.9$ is obtained. The moral of this approach is that we can still find a fitting
for a phenomenological formula encompassing both the anomalies prior to 2005 and the recent null results using only local parameters at perigee, but this comes at the cost of proposing a complicate dependence
on the parameter characterizing the orbital orientation (the inclination angle, $I$) and the parameters
corresponding to the location of the perigee (the latitude, $\phi$, and the altitude, $h$, at the point
of closest approach to the Earth).

If we assume that a new force field (a fifth force) is causing the anomalies, this would have a range of $R_E/\beta=291.2$ km but, instead of depending only of the distance, as the standard fifth force proposals \cite{Franklin,Fischbach}, it would also 
vary with latitude. So, our main conclusion in this paper is that, from a phenomenological point of view, there are good reasons
to propose that the flyby anomaly may be caused by a velocity-independent field of force generated by the rotating Earth decaying
very fast with the distance to the surface of the Earth and with a complex angular structure, still to be determined. This force
should depend on the mass of the object and it would be, fundamentally, gravitational in origin. It could also be connected with 
torsion or other extensions of General Relativity \cite{AcedoTorsion}.
But the only way to solve the riddle posed by the flyby anomaly is to obtain more
data as it could be provided by the ongoing Juno mission or future Earth flybys.

%
%

%

%
\section*{Acknowledgements}

I gratefully acknowledge the team behind the NASA Horizons' website for providing the ephemeris data
used in this paper.


%
 \bibliographystyle{plain}  
 \bibliography{acedobiblio}                

%

\end{document}